# A Study on Impact of Downsizing on Profitability of Construction Industries listed in Bombay Stock Exchange (BSE) India

D Reshma [1], Sudharani R [2*], Dr. Suresh N [3*]

[1]M.com Student, Faculty of Management and Commerce, Ramaiah University of Applied Sciences, Bengaluru, Karnataka, India – 560054.

Contact No.: +918317474413

[2]Assistant Professor, Faculty of Management and Commerce, Ramaiah University of Applied Sciences, Bengaluru, Karnataka, India – 560054

Contact No.: +918867658752

[3]Professor, Faculty of Management and Commerce, Ramaiah University of Applied Sciences, Bengaluru, Karnataka, India – 560054

Contact No.: +919980995377

[1]reshmadayalan4@gmail.com, [2]sudharani.co.mc@msruas.ac.in, [3]nsuresh.ms.mc@msruas.ac.in

***Abstract:*** *The study investigates the impact of downsizing/layoffs on profitability of construction industries listed in BSE India. In India construction industries have adopted downsizing long back in the organization to improve the firm's performance. For the purpose of the study, Secondary data of 15 Construction companies listed in BSE India have been considered for a period 10 years from FY.2010 to FY2019. Data has been taken from the company's official website. The variable considered for the analysis are Other Expenses, Returns on Net Worth, Employee Expenses, Number of Employees and Profit Per Employees. The study has used the Co-integration test to see co-integration between the variables, Ordinary Least Square (OLS) and Vector Auto Regression (VAR) model used for estimating the impact of downsizing on profitability of construction companies. OLS and VAR model has been used to drawn conclusion based on the P values and R square. Form the result it can be concluded that, Expect Profit Per Employees the downsizing variable which has no significant impact on the profitability of the firm's performance. Whereas the other Downsizing variables Employee Expenses and Number of Employee has a significant impact in the Profitability of the firm's performance.*

***Keywords:*** *Downsizing, Layoffs, Profitability, VAR, Other Expenses, Returns on net worth, Employee Expenses, Number of Employees, Profit Per Employees*

## 1. Introduction

Downsizing often takes place as part of a larger restructuring program at an organization. Downsizing is a strategy used to reduce the number of employees from the payroll, in order to improve the financial performance. Although it's usually thought of as a strategy organization use to become smaller. Downsizing can also be the result of company mergers, acquisition and takeovers, Its most common form comes in employee layoffs, which reduce payroll costs for the company. But the most common method or technique used by each organization are to ask the employees to leave the company. Employee downsizing is a planned set of organizational policies and practices aimed at workforce reduction with the goal of improving firm's performances. If too many companies start to cut payrolls, it can exacerbate the downturn and lead to higher unemployment. On the other hand, companies may downsize in order to improve their attractiveness to potential acquires, and their cost-cutting moves could result in a buyout offer. (Jesson Kenneth D' mello and Suresh N, 2019). Most importantly by downsizing it helps the organization. It was typically implemented during the economic downturns in order to improve efficiency and maintain profitability in organization. Thus, we view downsizing as an intentional event involving a range of organizational policies and actions undertaken to improve firm performance through a reduction in employees (Deepak K. Datta, 2010).

In India big companies such as L&T and other competing companies have undergone downsizing by removing all the employees who aren't needed. India's one of the biggest layoffs was in L&T where this step was an attempt to "right size" its strength in the face of the business slowdown. In India construction industry typically responds very sensitively to the economic, demographic, and political environment. Accordingly, for the long-term survival of a firm, growth through continuous innovation is very important. A focus on the construction sector brings into the picture certain key issues related to work conditions, recruitment patterns, migration, and cycles of exploitation (Verma Praveen, 2013).

But the construction industry is facing a lack of skilled labour too, thus this impact on the financial performance. The construction industry offers a lot of opportunity for employment but most of it is restricted to manual jobs. Due to the low wage expectancy of workers, who mainly come from the rural areas in search of work, contractors still follow traditional work practices. This doesn't just reduce the efficiency of the process but also makes it a lot risky. And around 70% contractors believe that there is lack of skilled workforce and qualified/certified professionals and sometimes this leads to downsizing in the company (Verma Praveen, 2013).

## 2. Literature Review

(Jesson Kenneth D'mello, and Suresh N (2019), The objective of this study is to determine the impact of layoff on the operations and profitability of the IT sector firms by looking at the service quality, technological levels, profitability and employee morale after the downsizing exercise. The study applied paired sampled t test has been done to know the impact of layoff and estimated the p value for conclusion. Fixed and random model has been used to drawn conclusion based on p values. The researcher concluded that, downsizing plays a significant role in the company's profitability. Laamanen, (2014) the researcher has done study based on correlation between the downsizing and firm's performance in an organizational regular perspective. Samples are been extracted from the organizations that are listed either on the STOXX Europe 50 or the Dow Jones Eurostoxx. A set of 73 firms in which total sections has 803 company, and computed with the help of GLS, cross-sectional time series regression, random effects and robustness test methods. The findings state that small scale downsizing leads to effectiveness improvements without distracting the existing routines and large-scale downsizing tends to be more valuable than medium scale downsizing. (Laamanen, (2014).

Udokwu, Ethel-Rose B, (2012) this study has made an objective to know the effect of downsizing on the laid off survivors' attitude towards the work in their bank. Researcher underwent primary data survey of 21 banks from Nigerian stock Exchange (NSE) and a population size of 2,304 workforce out of a sample of 341 workers were selected randomly. The researcher conclude that laid off survivors take downsizing as a wicked practice they feel less secured as their work is affected by downsizing practice. (Dominic H. Chai, (2010) As per this study examines the impact of foreign ownership on the firm's labour cost using a panel data of 496 publicly traded Korean. The objective of this study was to examine the relationship between economic and capital market pressures to disinvest in human capital. Hence, it shows that foreign ownership is positively related to labour cost but this positive effect is significantly weaker for firms with weak financial performance than those with strong financial performance (Dominic H. Chai, (2010).

Tim Goesaert, (2015) The researcher focused the short-term performance of downsizing firms. The present unique data set to study the short-term effects of downsizing on operational and financial performance of large German firms. obtained by examining 50,000 newspaper articles reporting on the 500 largest German firms. The results are robust against different ways to define the downsizing events, serial correlation, and a nonparametric approach to identify the impact of downsizing on efficiency, and especially downsizing for the latter firms appears to be unsuccessful. (Paul williams, (2011) the study focused on impact of downsizing on customer satisfaction through the survivors and what impact it has and how it will create on the financial performance of the firm. A very less research has inspected the effect of downsizing on customers the researcher has taken a step via a case study of a Fortune 100 company before a noteworthy downsizing occasion and 994 customers later. Downsizing impacts on 3 different ways to customer satisfaction such as the attitude of the survivors in the supply chain specially in business to business system. Therefore, researcher concluded downsizing has a direct negative impact on customer satisfaction levels and on anticipated retention rates which will escort to a direct negative financial below on the services contributor due to loss of expected upcoming consumer profits (Paul williams, (2011).

Researcher explores the effect of downsizing on both a firm's financial performance in terms of profitability and efficiency, and a firm's employee productivity. The study has been analyzed six-year longitudinal financial data of 258 listed from Inc.-Financial Analysis System (KIS-FAS) database. The study used multiple regression, the paper investigated the relationship between downsizing and three measures of financial performance and two measures of organizational performance. Therefore, after comparing these five performance data during downsizing periods and after downsizing periods, it was clear that downsizers had relatively lower mean scores on most variables than non-downsizers (Gyu Chang Yu, (2004).

## 3. Materials and Methods

### 3.1: Data description and methodology

The objective of this study is to develop a model, to know impact of the downsizing variables on the profitability of the firm's performance. This study has used the secondary data for analysis. The analysis of 15 Construction companies' data are collected from the Money Control websites BSE India.

- The data of 10 years have been considered for the analysis from the period of FY2010 to FY2019. The variables considered for the analysis are Other expenses, Returns on net worth, Employee Expenses, Number of Employees, Profit Per Employees
- Panel unit root test is conducted to test if the time series variables are stationary or non-stationary and possess unit root
- By estimating equation, Ordinary least square model is generated for the variables collected by considering profitability as the dependent variables and downsizing as the independent variables
- By estimating equation two different models has been developed to estimate a better model
- The variables which have probability less than 5% are impacting on profitability of the firm's performance
- Co-integration test is developed to check whether there is co-integration within the selected variables
- Least square unrestricted Vector Auto Regression (VAR) model was developed to check the long run effect of variables on the profitability
- Wald test was generated to check the short run effect of variables on the profitability

Specification of the estimated model equation is explained here:

**Model 1: OE** $= \beta + \beta_1 EE + \beta_2 NOE + \beta_3 PPE$ (1)
**Model 2: RONW** $= \beta + \beta_1 EE + \beta_2 NOE + \beta_3 PPE$ (2)

Where, Y1 OE is Other Expenses and Y2 RONW is Returns on Net Worth which is dependent variables.

**β** = Constant
**EE** = Employee Expenses
**NOE** = Number of Employees
**PPE** = Profit Per Employees

Finally, the best model will be developed for the variable which has impact of downsizing on profitability from the 2 models.

## 4. Results and discussion

This study presents the results of the analyzed secondary data collected for the study. This study has features of the data set along with the statistical package used for analysis. In this analysis, VAR has been used to examine the profitability respond to different downsizing variables. If there are any concerns surrounding the endogeneity of the variables "EVIEWS" provides a frame work to explore the iterations as well as variables that are considered endogenous and allowed to impact on the other system variable.

Table 1: Descriptive Statistics

| Variables | OE | RONW | EE | NOE | PPE |
|---|---|---|---|---|---|
| Mean | 73.1005 | 3.0559 | 63.1144 | 3374.85 | 0.21829 |
| Median | 53.015 | 4.2755 | 49.012 | 927 | 0.205 |
| Minimum | -98.822 | -106.608 | -146.318 | -945.2 | -3.4956 |
| Maximum | 448.79 | 28.23 | 302.56 | 50544.3 | 3.786 |
| SD | 86.5015 | 20.47561141 | 81.4622 | 8678.989648 | 0.667 |

**Source:** Based on Eviews output, Authors' Analysis

From the table 1 results obtained is a descriptive statistic for the selected variables. In the above 10 years of data has been taken for the analysis for the period FT2010 to FT2019. The minimum Other Expenses was -98.822 in 2010 and the maximum was 448.79 in 2019 and the minimum for Returns on net worth was -106.608 in 2010 and the maximum was 28.23 in 2019. There has continuous increase in the in the Other Expenses and in Returns on net worth from the stage of downsizing.

For variable prospective regression model (Estimating equation- ordinary least square method) has been built to understand the impact of downsizing variables on the profitability. Here, dependent variable is OE and RONW and the independent variables are as follows: Employee Expenses, Number of Employees, Profit Per Employees (Table 2) and (Table 3).

Table 2: Estimating equation ordinary least square method

| Dependent Variable – OE (ordinary least square method) | | | | |
|---|---|---|---|---|
| Variables | Coefficient | Std. Error | t-statistics | Prob. |
| C | -32.38845 | 8.227573 | -3.936575 | 0.0001 |
| EE | 0.593009 | 0.076125 | 7.789931 | 0.0000 |
| NOE | 0.002061 | 0.000698 | 2.952239 | 0.0037 |
| PPE | 16.81683 | 9.172648 | 1.833367 | 0.0688 |
| $R^2$ 0.853626 | F-statistic 51.21917 | | Prob(F-statistic) 0.000000 | |
| Akaike info criterion 6.442977 | | | Schwarz criterion 6.523261 | |

**Source:** Based on Eviews output, Authors' Analysis

Table 3: Estimating equation ordinary least square method

| Dependent Variable – RONW (ordinary least square method) | | | | |
|---|---|---|---|---|
| Variables | Coefficient | Std. Error | t-statistics | Prob. |
| C | -5.913454 | 2.069402 | -2.857567 | 0.0049 |
| EE | 0.098935 | 0.019147 | 5.167123 | 0.0000 |
| NOE | 0.000577 | 0.000176 | 3.284589 | 0.0013 |
| PPE | 3.567317 | 2.307108 | 1.546229 | 0.0042 |
| $R^2$ 0.823745 | F-statistics 53.2302 | | Prob(F-statistics) 0.000000 | |
| Akaike info criterion 7.682514 | | | Schwarz criterion 7.762796 | |

**Source:** Based on Eviews output, Authors' Analysis

From the above model obtained we can see that in table 1 OE the $R^2$-value is 0.853626 and the probability is 0.0000 and in table 2 RONW the $R^2$-value is 0.823745 and the probability is 0.0000 which represents the best part of the model. From the above model, we have removed the variable average business per employees because the probability was showing more than 5% which says that it has no impact on Profitability. The most important condition for selecting the best model is values of Akaike info criterion and Schwarz criterion should be less. So, here we can proceed with this model where from both the model are showing that all the independent variable has p value less than 5%. From the above analysis, we came to know which economic variables are affecting the profitability of the firm's performance.

### 4.1 Panel Unit Root Test:

Panel unit root test has been done to know the variables are stationary or non-stationary. After examining stationary of the data with data with the Levin, Lin and Chu test, Breitung Im, Pesaran and Shin, Fisher type using ADF and PP test. It can be stated that all our variables are stationary at level and at 1$^{st}$ difference they become stationary. The test was executed with a significance level of 5%.

### 4.2 Panel Co-integration Test:

Panel co-integration test has been conducted to check whether there is co-integration within the selected variables. This test proved that there is a co-integration within the selected variables. So, to prove there is a co-integration between the variables the Vector Auto Regression (VAR) is conducted for the further analysis to check the long run effect.

### 4.3 Vector Auto Regression (VAR):

It is an econometric model used to identify the linear interdependencies among different time series. It helps to identify whether the impact of variable on another variable is in long run or in short run (Prajwal P.P and Suresh N, 2017).

Lag test, Lag 1 and 2 has been conducted on 2 dependent variable OE and RONW. The selected lag OE which is distributed between lags 3 and 4, but mostly on lag order 4. The main condition is that Akaike info criterion should be less this condition applied because the lower the value, the better the model. And for OE the Akaike info criterion is 8.944777 which is less than other lags. i.e., in lag 1 is 9.102246, lag 2 is 9.124304, lag 3 is 9.138051 and for the other dependent variable RONW the Lag which is distributed between lags 1 and 2, but mostly on lag order 2 the Akaike info criterion is 6.565155 which is less than other lags. So, in lag 1 is 6.649744, lag 3 is 6.5866430, lag 4 is 6.597838. We can conclude that the optimal lag length for the model 1 is 4 and model 2 is 2 the best criterion to adopt for the model is Akaike info criterion. Hence, the Lag 4 and Lag 2 needs to be selected for creating the VAR model.

- **H₀:** OE has no significant impact of downsizing on the profitability of the company
- **H₁:** OE has a significant impact of downsizing on the profitability of the company

Table 4: Estimating Vector Auto Regression model

| Dependent variable- OE (Panel Least Square method (Unrestricted VAR)) ||||
|---|---|---|---|---|
| Variables | Coefficient | Std. Error | t-statistics | Prob. |
| C(1) | 0.833269 | 0.121706 | 6.846583 | 0.0000 |
| C(2) | 0.252779 | 0.172114 | 1.468669 | 0.1457 |
| C(3) | 0.540832 | 0.195376 | 2.768153 | 0.0070 |
| C(4) | -0.642783 | 0.144750 | -4.440628 | 0.0000 |
| C(5) | -0.974328 | 3.797298 | -0.256584 | 0.7981 |
| C(6) | 0.043602 | 0.035914 | 1.214052 | 0.0282 |
| C(7) | 0.000127 | 0.000366 | 0.347833 | 0.0089 |
| C(8) | 4.126245 | 4.683382 | 0.881040 | 0.3809 |
| R² 0.951755 | F-statistics 231.093 || Prob(F-statistics) 0.000000 ||
| Akaike info criterion 8.944777 ||| Schwarz criterion 9.166982 ||

**Source:** Based on Eviews output, Authors' Analysis

**Estimated equation used for the analysis**

OE= C(1)*OE(-1) + C(2)*OE(-2) + C(3)*OE(-3) + C(4)*OE(-4) + C(5) + C(6)*EE + C(7)*NOE + C(8)*PPE                                                                                                                             (3)

In the Table 4 result which is obtained through the vector auto regression model. This is formed from the above stated equation in order to see whether these variables have a long run effect on the dependent variable which is OE. In the above table C(4) is the dependent variable (OE) and C(6), C(7), C(8) is independent variables. We can see that the probability value is less than 0.05 in EE and NOE which says that there is long-term effect of independent variables on dependent variables. But in PPE probability value is more than 0.05 which states that there is no long-term effect in independent variables on dependent variables.

**Wald Test on OE:**

- **H₀:** OE has no short-term effect between the independent variables
- **H₁:** OE has a short-term effect between the independent variables

**Table 5: Wald Test 1 on OE**

| Test statistic | Values | df | Probability |
|---|---|---|---|
| t-statistic | 1.214052 | 82 | 0.0282 |
| F-statistic | 1.473921 | (1,82) | 0.0282 |
| Chi-square | 1.473921 | 1 | 0.0047 |

Null Hypothesis: C(6)=0

**Table 6: Wald Test 2 on OE**

| Test statistic | Values | df | Probability |
|---|---|---|---|
| t-statistic | 0.347833 | 82 | 0.0089 |
| F-statistic | 0.120988 | (1, 82) | 0.0089 |
| Chi-square | 0.120988 | 1 | 0.008 |

Null Hypothesis: C(7)=0

**Table 7: Wald Test 3 on OE**

| Test statistic | Values | df | Probability |
|---|---|---|---|
| t-statistic | 0.88104 | 82 | 0.3809 |
| F-statistic | 0.776231 | (1, 82) | 0.3809 |
| Chi-square | 0.776231 | 1 | 0.3783 |

Null Hypothesis: C(8)=0

**Source:** Based on Eviews output, Authors' Analysis

In the Table 7 it shows that profitability value of C(8) Profit per employee is (0.3809) which shows a long term effect which states that there is no significant effect on the profitability of the company but in C(6) Employee expense (0.0282), C(7) Number of employee (0.0089) which shows that there is a short term effect between independent variable on dependent variable and proves that VAR model has long term effect except PPE which shows that there is a long-term impact on downsizing of profitability of construction industries.

- **H₀:** RONW has no significant impact of downsizing on the profitability of the company
- **H₁:** RONW has a significant impact of downsizing on the profitability of the company

**Table 8: Estimating Vector Auto Regression model**

| Dependent variable- RONW (Panel Least Square method (Unrestricted VAR)) | | | |
|---|---|---|---|
| Variables | Coefficient | Std. Error | t-statistics | Prob. |
| C(1) | 0.445973 | 0.109569 | 4.070231 | 0.0001 |
| C(2) | 0.336534 | 0.125223 | 2.687480 | 0.0087 |
| C(3) | 0.047422 | 0.124573 | 0.380673 | 0.7044 |
| C(4) | -0.100945 | 0.106906 | -0.944240 | 0.0478 |
| C(5) | 1.682149 | 1.329849 | 1.264917 | 0.2095 |
| C(6) | 0.013549 | 0.009072 | 1.493616 | 0.0391 |
| C(7) | 0.000158 | 0.000108 | 1.465183 | 0.0467 |
| C(8) | 0.561821 | 1.425930 | 0.394003 | 0.6946 |
| $R^2$ 0.810597 | F-statistics 50.13430 | | Prob(F-statistics) 0.000000 |
| Akaike info criterion 6.565155 | | | Schwarz criterion 6.820043 |

**Source:** Based on Eviews output, Authors' Analysis

**Estimated equation used for the analysis**

RONW= C(1)*RONW(-1) + C(2)*RONW(-2) + C(3)*RONW(-3) + C(4)*RONW(-4) + C(5) + C(6)*EE + C(7)*NOE + C(8)*PPE                                                                                                                          (4)

In the Table 8 result which is obtained through the vector auto regression model. In the above table C(4) is the dependent variable (RONW) and C(6), C(7), C(8) is independent variables. We can see that the probability value is less than 0.05 in EE and NOE which says that there is long-term effect of independent variables on dependent variables. But in PPE probability value is more than 0.05% which states that there is no long-term effect in independent variables on dependent variables.

**Wald Test on RONW**

- **H₀:** RONW has no short-term effect between the independent variables
- **H₁:** RONW has a short-term effect between the independent variables

**Table 9: Wald Test 1 RONW**

| Test statistic | Values | df | Probability |
|---|---|---|---|
| t-statistic | 1.493616 | 82 | 0.0391 |
| F-statistic | 2.230888 | (1, 82) | 0.0391 |
| Chi-square | 2.230888 | 1 | 0.0352 |

Null Hypothesis: C(6)=0

**Table 10: Wald Test 2 RONW**

| Test statistic | Values | df | Probability |
|---|---|---|---|
| t-statistic | 1.465283 | 82 | 0.0467 |
| F-statistic | 2.146762 | (1, 82) | 0.0467 |
| Chi-square | 2.146762 | 1 | 0.0429 |

Null Hypothesis: C(7)=0

**Table 11: Wald Test 3 RONW**

| Test statistic | Values | df | Probability |
|---|---|---|---|
| t-statistic | 0.394003 | 82 | 0.6946 |
| F-statistic | 0.155239 | (1, 82) | 0.6946 |
| Chi-square | 0.155239 | 1 | 0.6936 |

Null Hypothesis: C(8)=0

**Source:** Based on Eviews output, Authors' Analysis

In the Table 11 shows that probability value of C(8) Profit per employee is (0.6946) which shows a long term effect which states that there is no significant effect on the profitability of the company but in C(6) Employee expense (0.0391), C(7) Number of employee (0.0467) which shows that there is a short term effect between independent variable on dependent variable and proves that VAR model has long term effect except PPE which shows that there is a long-term impact on downsizing of profitability of construction industries.

# 5. Conclusion

Overall, the study has commenced an attempt to discuss the theory of downsizing. The study deals with the state of downsizing in Indian context. And the relationship between the downsizing and financial performance of construction industries in India was explored in this study. The panel unit root test was done where it concludes that there is a significant impact in the variables and difference in financial performance of construction industries in India. The panel analysis used in this paper is to explain the impact of downsizing and relationship between the profitability of the construction companies. The co-integration test was done where the result stated that there is a cointegration between variables. The results show that some of the predominant factors affecting the profitability are Number of Employees, and Employee expenses which are the one of the most effecting downsizing variables expect the Profit Per Employees in both the model OE and in RONW. Based on the results the study also highlighted that the downsizing relationship between the profitability were found to be significantly affecting the organization that went for downsizing. Lack of qualified and certified professional were also an important reason which forces the organization for reduction of employees. The result even stated that two variables are showing the impact on the OE and RONW which showed that company can actually be stable and improve the efficiency after downsizing.

## 6. Limitation and Suggestions for the future research

The study was within a short study period which may not have allowed the researcher to consider more variables and to measure more phenomenon related to the study variables. For the further analysis more variables can be added. And the study was only limited to 15 companies for the period of 10 years data from FY.2010- FY.2019. For future research the downsizing in the company should be informed earlier to the employees. So, that layoffs survivors don't face downsizing as unjust and unfair practice, without informed they feel guilt and sympathetic for laid-off victims, and they feel less secured (Udokwu, Ethel Rose B 2012). Therefore, a similar study can be made for other industries listed in Indian National Stock Exchange. There is scope for conducting a study to analyze the effect of downsizing on profitability of other companies, especially financial companies, Different sectors, service companies and Non listed companies. And further study can also be done on important issues related to downsizing.